\documentclass[twocolumn,secnumarabic,amssymb, nobibnotes, aps, prl, superscriptaddress]{revtex4-2}

\usepackage{graphicx}
\usepackage{dcolumn}
\usepackage{bm}
\usepackage{amsmath}
\usepackage{enumerate}
\usepackage{setspace}
\usepackage{dsfont}
\usepackage{subfigure}
\usepackage{multirow}
\usepackage{indentfirst} 
\usepackage {mathrsfs}
\usepackage{color}
\usepackage{upgreek}

\usepackage{threeparttable}  
\usepackage[pdfstartview=FitH,
CJKbookmarks=true,
bookmarksnumbered=true,
bookmarksopen=true,
colorlinks, 
pdfborder=001,
linkcolor=blue,
anchorcolor=blue,
citecolor=blue
]{hyperref}

\begin{document}

\title{Node-downloadable frequency transfer system based on a mode-locked laser with over 100 km of fiber}%

\author{Ziyi Jin}%
\affiliation{State Key Laboratory of Information Photonics and Optical Communications, Beijing University of Posts and Telecommunications, Beijing 100876, China}

\author{Ziyang Chen}%
\email[E-mail: ]{chenziyang@pku.edu.cn}
\affiliation{State Key Laboratory of Advanced Optical Communication Systems and Networks, School of Electronics, and Center for Quantum Information Technology, Peking University, Beijing 100871, China}

\author{Kai Wu}%
\affiliation{State Key Laboratory of Information Photonics and Optical Communications, Beijing University of Posts and Telecommunications, Beijing 100876, China}

\author{Dongrui Yu}%
\affiliation{State Key Laboratory of Advanced Optical Communication Systems and Networks, School of Electronics, and Center for Quantum Information Technology, Peking University, Beijing 100871, China}

\author{Guohua Wu}
\affiliation{State Key Laboratory of Information Photonics and Optical Communications, Beijing University of Posts and Telecommunications, Beijing 100876, China}

\author{Song Yu}
\affiliation{State Key Laboratory of Information Photonics and Optical Communications, Beijing University of Posts and Telecommunications, Beijing 100876, China}

\author{Bin Luo}
\affiliation{State Key Laboratory of Information Photonics and Optical Communications, Beijing University of Posts and Telecommunications, Beijing 100876, China}

\author{Hong Guo}
\email[E-mail: ]{hongguo@pku.edu.cn}
\affiliation{State Key Laboratory of Advanced Optical Communication Systems and Networks, School of Electronics, and Center for Quantum Information Technology, Peking University, Beijing 100871, China}

\date{\today}%

\begin{abstract}
To meet the requirements of time-frequency networks and enable frequency downloadability for nodes along the link, we demonstrated the extraction of stable frequency signals at nodes using a mode-locked laser under the condition of 100 km laboratory fiber. The node consists of a simple structure that utilizes widely used optoelectronic devices and enables plug-and-play applications. In addition, the node can recover frequency signals with multiple frequencies, which are useful for scenarios that require different frequencies. Here, we experimentally demonstrated a short-term frequency instability of $2.83\times {{10}^{-13}}$@1 s and a long-term frequency instability of $1.18\times {{10}^{-15}}$@10,000 s at the node, which is similar to that at the remote site of the frequency transfer system. At the same time, frequency signals with different frequencies also achieved stable extraction with the same performance at the node. Our results can support the distributed application under large-scale time-frequency networks.
\end{abstract}

\maketitle

\section{Introduction}
With the rapid development of atomic clocks, a large variety of applications require precise time-frequency standards such as geodesy, fundamental physics, and other areas\cite{mcgrew2018atomic, grotti2018geodesy, cliche2006precision, schiller2013feasibility, kolkowitz2016gravitational, chou2010optical, delva2017test, derevianko2014hunting, stadnik2015searching}. The construction of a high-precision multi-user time-frequency transfer network is an important issue. In recent years, optical time-frequency transfer technology has developed rapidly\cite{krehlik2015frequency,gao2023comparison, predehl2012920,lisdat2016clock,riehle2017optical,schioppo2022comparing,gozzard2022ultrastable, marra2010high,hou2011long,giorgetta2013optical,collaboration2021frequency,kang2019free,shen2022free, caldwell2023quantum, droste2013optical, qian2017precise, wang2018stable, wang2020fiber, tian2020hybrid}. Time-frequency transfer based on modulated-laser technology has been tested at a distance of 3,000 km\cite{krehlik2015frequency,gao2023comparison}. Through optical frequency transfer technology, the synchronization ability of high-precision optical clock networks can be realized\cite{predehl2012920, lisdat2016clock,riehle2017optical,schioppo2022comparing,gozzard2022ultrastable}. It is worth mentioning that the time-frequency transfer system based on optical frequency comb (OFC) has also reached a new stage that opens a new pathway for time and frequency transfer\cite{marra2010high,giorgetta2013optical,collaboration2021frequency,kang2019free,shen2022free, caldwell2023quantum, hou2011long, yang201710}.

As an excellent transmission medium for optical waves, fiber has the advantages of high bandwidth, low attenuation, and anti-interference. Fiber-based time and frequency transfer has been widely demonstrated to be one of the mature technologies for long-distance time and frequency transfer\cite{krehlik2015frequency,gao2023comparison, predehl2012920, lisdat2016clock, schioppo2022comparing, marra2010high, droste2013optical, qian2017precise, wang2018stable, wang2020fiber, tian2020hybrid}. Fiber-based time-frequency transfer technology has the potential to facilitate the establishment of time-frequency transfer networks efficiently. In the past few decades, various multi-node time-frequency network schemes have been proposed in addition to various point-to-point technical systems\cite{grosche2008verfahren, gao2012fiber, krehlik2013multipoint, sliwczynski2015multipoint, schediwy2013high, bai2015fiber,wang2015square,jiang2015precise,liu2016gvd,cui2017passive,shang2018stable,10092931,gao2023multi}. Grosche \emph{et al.} proposed a method that can obtain a frequency at any node in the optical fiber link of the time-frequency transfer system\cite{grosche2008verfahren}. This method can form a bus topology time-frequency transfer network. Gao \emph{et al.}\cite{gao2012fiber} and Krehlik \emph{et al.}\cite{krehlik2013multipoint} also proposed similar methods. Sliwczynski \emph{et al.} improved the above method so that not only nodes but also side branches can be added to the main chain to form a tree topology network\cite{sliwczynski2015multipoint}. Schediwy \emph{et al.} also proposed a multi-address transmission scheme that can form a star topology type network\cite{schediwy2013high}. Besides the scheme above, numerous analogous works have been accomplished and have achieved favorable outcomes\cite{bai2015fiber,wang2015square,jiang2015precise,liu2016gvd,cui2017passive,shang2018stable,10092931,gao2023multi}. Practical technical support and research foundation for the construction of large-scale time-frequency networks are provided by these technologies.

However, in order to achieve efficient node downloadable technology, there are usually two considerations at the node. First is the complexity of the node system, and second is using different frequency signals at the node. Although the above scheme has many advantages in accuracy and scalability, the node system's complexity and the reference signal frequency limitation can be further optimized.

In this paper, in order to simultaneously obtain different frequency signals at the node and simplify the node structure, we propose and experimentally verify a plug-and-play node-downloadable frequency transfer scheme based on mode-locked-laser (MLL) technology that allows recovering frequency signals at nodes. To achieve this goal, the light source used in this scheme is an MLL source that is locked to the clock signal through multi-collaborative feedback control locking technology to resist rapid and drastic environmental changes. A high-precision direct digital synthesizer (DDS) compensates for the phase drift introduced during transmission. The node fully utilizes the multi-harmonic characteristics of the MLL by using common experimental devices such as filters and mixers, which can download different frequency harmonic components with high stability without additional active compensation devices. This experiment achieved frequency transfer with frequency instabilities of $2.79\times {{10}^{-13}}$@1 s and $1.25\times {{10}^{-15}}$@10,000 s under 100 km laboratory fiber conditions. At the same time, we recovered frequency signals at the node (about 50 km from the local site) in this experimental structure. Multiple harmonic components like 100 MHz, 200 MHz, and 400 MHz can yield effective output. The frequency instabilities can reach $2.83\times {{10}^{-13}}$@1 s and $1.18\times {{10}^{-15}}$@10,000 s, which are similar to that at the remote site, thus proving the effectiveness of this scheme. Our experiment demonstrates the efficiency of MLL technology in achieving efficient multi-frequency downloading at nodes. It also provides technical support for the use of frequency signals at intermediate nodes in future time-frequency networks.

This paper is organized as follows: In Section 2\ref{2}, we introduce the experimental setup and the principle of this experiment, including the light source (Subsection 2.1\ref{2.1}), collaborative feedback control system (Subsection 2.2\ref{2.2}), active pre-compensation system (Subsection 2.3\ref{2.3}) and DWDM-based node downloading (Subsection 2.4\ref{2.4}). In Section 3\ref{3}, we show the experimental results and analyses, including the performance test of the collaborative feedback control system (Subsection 3.1\ref{3.1}), the active pre-compensation system (Subsection 3.2\ref{3.2}) and DWDM-based node downloading (Subsection 3.3\ref{3.3}). Finally, we summarize our work in Section 4\ref{4}.

\section{Experimental setup.} 
\label{2}
\begin{figure*}[ht!]
\centering\includegraphics[width=12cm]{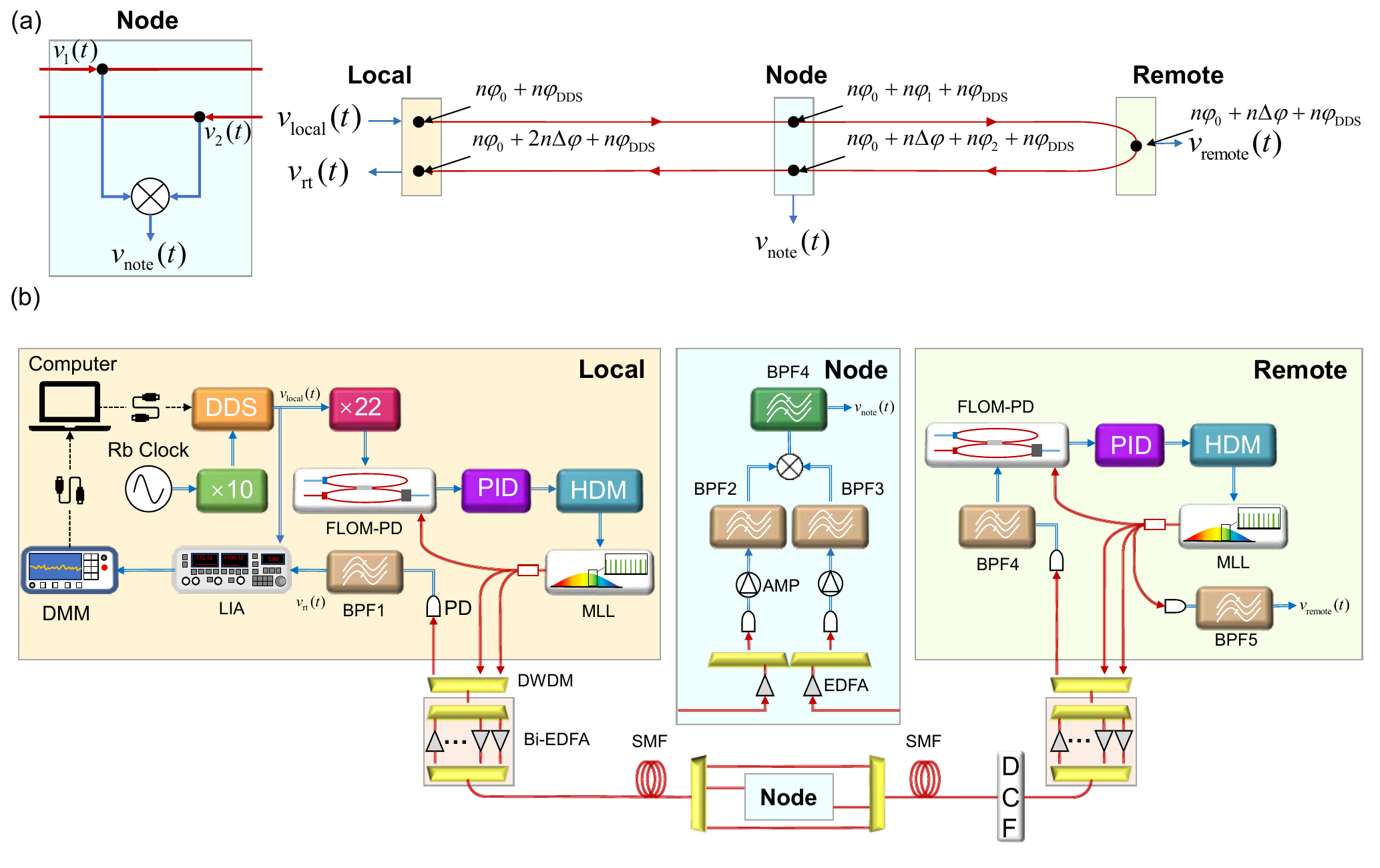}
\caption{Node-downloadable experimental setup based on MLL. (a) Simplified block diagram of node-downloadable frequency transfer system; (b) Detailed block diagram of node-downloadable frequency transfer system. MLL: mode-locked laser with repetition frequency locked by a collaborative feedback control system; DWDM: dense wavelength division multiplexer; PD: photoelectric detector; EDFA: erbium-doped fiber amplifier; Bi-EDFA: bidirectional-erbium-doped fiber amplifier; BPF4: band-pass filter with a center frequency of 200 MHz and bandwidth of 10 MHz; BPF1, BPF2, BPF3, BPF5 and BPF6: band-pass filters with a center frequency of 100 MHz and bandwidth of 10 MHz; FLOM-PD\cite{jung2012subfemtosecond}: fiber-loop optical-microwave phase detector; PID: proportional integral derivative control system; DMM is 6½ digit digital multimeter; AMP: electrical amplifier; LIA: lock-in amplifier; HDM: high-voltage driver module, which is responsible for amplifying the control signal output by PID and providing load voltage for the PZT inside the MLL; DCF: dispersion compensation fiber.}
\label{Experimental}
\end{figure*}

The node downloadable experimental setup based on MLL is shown in FIG. \ref{Experimental}. We use a homemade MLL with a repetition frequency (${f_{\rm r}}$) of 100 MHz as the light source, with a central wavelength of about 1,560 nm, a spectral width of about 10 nm, and an optical power of about 100 mW. The MLL's ${f_{\rm r}}$ is controlled by the collaborative feedback control system, which uses a piezoelectric ceramic transducer (PZT), temperature control (TEC), and delay line. The local site and the remote site are connected by two coils of 50 km wound single-mode fiber (SMF) and a dispersion compensation fiber (DCF) with a length of about 10 km, where the length of the dispersion compensation fiber is obtained after strict dispersion calculation. The total loss of SMF is 21.12 dB, while the total loss of DCF is 12.46 dB. Since the spectrum of the MLL pulse is wide, it completely covers the ITU standard channels of the dense wavelength division multiplexer (DWDM) we use, namely C30-C37. To minimize the crosstalk between different wavelength channels and affect the frequency instability of the system, we choose four channels C30, C31, C36, and C37 for the experiment, with a spectral width of 0.8 nm (100 GHz). Among these, the optical signals in channels C30 and C31 are transmitted forward, and those in channels C36 and C37 are transmitted backward. To ensure the optimal system performance under current circumstances, it is imperative to maintain rigorous control over the incoming power of the optical fiber at both the local and remote sites \cite{jin2022analyzing}. Especially as the number of nodes increases, a thorough analysis is necessary to determine the most effective power control measures. We will leave a more detailed analysis of this issue for future research.

 FIG. \ref{Experimental} and \ref{CFCC} show the specific experimental structure, which consists of a collaborative feedback control system, active pre-compensation system, node downloading, and other components.

\subsection{Light source.} \label{2.1} At the local site, to achieve high stability locking of the ${f_{\rm r}}$, and thus complete high-performance microwave frequency transfer, this experiment first locks the MLL as the carrier on the microwave clock. We employ a feedback control system to achieve high-accuracy and long-duration ${f_{\rm r}}$ locking of the homemade MLL. The feedback control system consists of PZT, TEC, and delay line, which can adjust ${f_{\rm r}}$ to different extents. The delay line has an extensive adjustable range and fast adjustment speed but slightly lower precision. Its minimum step value is 3 $\upmu$m, and the adjustable range is 0-39,999 $\upmu$m. The corresponding minimum step value of ${f_{\rm r}}$ adjustment is 95 Hz, and the related ${f_{\rm r}}$ adjustable range is 0-3.80 MHz. PZT's adjustable range is the smallest, with the fastest adjustment speed and the highest precision. Its setting range is 0-120 Hz, and the bandwidth is around 10 kHz. TEC can only achieve slow adjustment of the cavity length, with a precision of 0.001 ℃, a tuning range of ±3.91 ℃, and a corresponding ${f_{\rm r}}$ adjustment range of 6.57 kHz.

\subsection{Collaborative feedback control system.}\label{2.2}
To enable the use of this scheme in environments with drastic temperature changes, we use computer equipment to control the delay line, TEC, and PZT in real-time. When the ambient temperature of the MLL changes significantly, the high-precision and long-term ${f_{\rm r}}$ locking of the MLL can be achieved to meet the requirements of the subsequent experiment.

\begin{figure*}[ht!]
\centering\includegraphics[width=12cm]{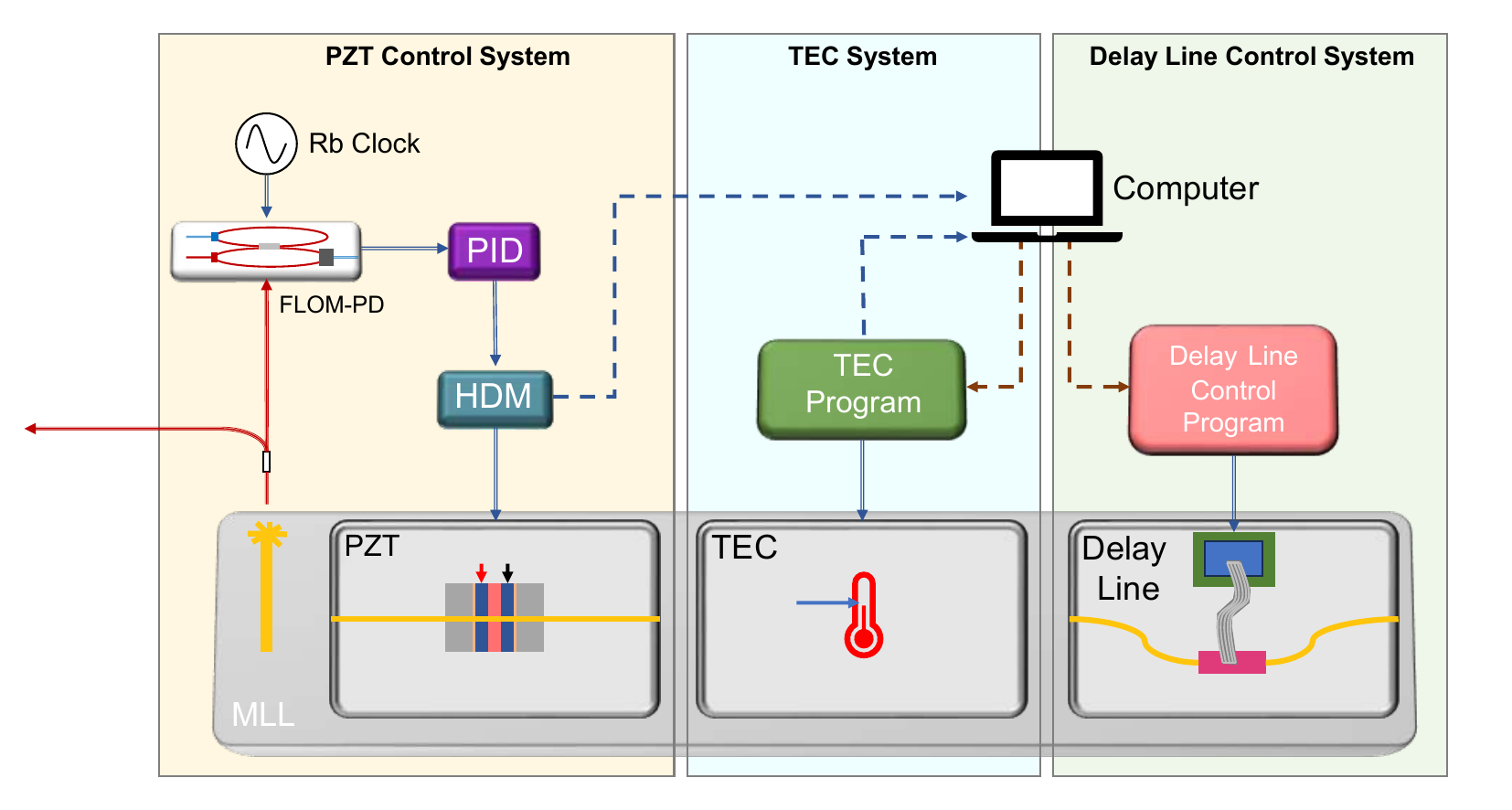}
\caption{Schematic diagram of the locking system of the MLL with collaborative feedback control of PZT, TEC, and delay line. MLL: mode-locked laser; PZT: piezoelectric ceramic transducer; TEC: temperature control; PID: proportional integral derivative control system; HDM: high-voltage driver module.}
\label{CFCC} 
\end{figure*}

The working principle of the collaborative feedback control system is described as follows: The collaborative feedback control system is composed of three sub-servo stabilization systems: the PZT control system, the TEC system, and the delay line control system, as illustrated in FIG. \ref{CFCC}. While the PZT control system is working, the TEC system will monitor the current driving voltage of PZT, which is used to regulate the temperature of the fiber in the MLL. The TEC system will calculate the temperature adjustment amount of the fiber in the resonant cavity of the MLL according to the current driving voltage value of PZT and then adjust the TEC in it to control its temperature setting value. When the PZT driving voltage deviates from the set value, TEC will adjust the temperature of the fiber in the resonant cavity according to the fiber temperature adjustment amount so that the PZT driving voltage is maintained near the set value. In most cases, the delay line control system is in a standby state and will monitor the current driving voltage of PZT and the output voltage of TEC. There are two possible scenarios in the experiment: 1) the difference between the set fiber temperature value and the current fiber temperature value is too large, exceeding the tuning range of TEC, and TEC cannot adjust the temperature even at full load. This is indicated by the high output voltage value of TEC; 2) the ambient temperature changes too rapidly, and the temperature tuning speed is too slow. This is indicated by a large deviation of the current driving voltage of PZT from the set value. When these two situations occur, the delay line control system activates, adjusting the fiber delay line in the MLL, directly changing the MLL cavity length, and making the PZT driving voltage and TEC setting temperature return to normal range. In order to improve the real-time processing speed of the system, we use a computer to coordinate the control of the delay line, temperature control, and PZT and achieve long-term locking of MLL in an environment with severe temperature fluctuations.

\subsection{Active pre-compensation system.}\label{2.3}
Active pre-compensation is used in the experimental device to eliminate the phase drift introduced by the signal transmission, and the specific principle is explained in detail below. The signal source synchronized with the frequency signal output from the Rb clock outputs a reference signal ${{v}_{0}}(t)$ with a frequency of 100 MHz, and for simplicity, the following signal expressions all ignore the influence of amplitude. ${{v}_{0}}(t)$ can be expressed as:

\begin{equation} \label{eq1}
{{v}_{0}}(t)=\cos ({{\omega }_{0}}t+{{\varphi }_{0}}),
\end{equation}
where ${{\omega }_{0}}=100\text{ MHz}$, and ${{\varphi }_{0}}$ represents the initial phase of the signal.

The signal serves as the reference signal for the DDS, which produces a 100 MHz signal that can be represented as
\begin{equation} \label{eq2}
{{v}_{\text{local}}}(t)=\cos ({{\omega }_{0}}t+{{\varphi }_{0}}+{{\varphi }_{\text{DDS}}}).
\end{equation}

A power splitter divides the ${v}_{\text{local}}(t)$, with one being the reference signal for the lock-in amplifier (LIA) and the other being transformed by a frequency doubler into a 2.2 GHz signal, which is used as the reference for the fiber-loop optical-microwave phase detector (FLOM-PD)\cite{jung2012subfemtosecond}. The signal output from the local MLL passes through a $1\times 3$ fiber coupler, and the first optical signal is attenuated to about 10 dBm by an attenuator and then connected to the local FLOM-PD for phase detection with the reference signal. The error signal obtained by phase detection is used for the collaborative feedback control system to lock ${f_{\rm r}}$. The other two optical signals enter the channels C30 and C31 of the local DWDM, respectively, and then the optical fiber link for forward transmission. The \textit{n}-th harmonic frequency component of the optical signal in the channel C30 is denoted as signal ${{v}_{\text{C30}}}(t)$, and that in the channel C31 is denoted as ${{v}_{\text{C31}}}(t)$, which can be expressed as
\begin{equation} \label{eq3}
\begin{split}
{{v}_{\text{C30}}}(t)=\cos (n{{\omega }_{0}}t+n{{\varphi }_{0}}+n{{\varphi }_{\text{DDS}}}), \\
{{v}_{\text{C31}}}(t)=\cos (n{{\omega }_{0}}t+n{{\varphi }_{0}}+n{{\varphi }_{\text{DDS}}}), 
\end{split}
\end{equation}
where $n=1,2,3...$ The different channels do not affect the phase and frequency of their frequency components.

The signal transmitted forward by the local DWDM passes through the local bidirectional-erbium-doped fiber amplifier (Bi-EDFA) and is amplified to about 3 dBm before being connected to the optical fiber link. At the remote site, the remote Bi-EDFA amplifies the optical signal leaving the fiber link to about 3 dBm. Then, the amplified optical signal is connected to the remote DWDM. The optical signal in channel C30 is converted by a photoelectric detector (PD), and then the first harmonic frequency component ${{v}_{\text{remote}}}(t)$ is filtered out by a band-pass filter (BPF) with a center frequency of 100 MHz. ${{v}_{\text{remote}}}(t)$ can be expressed as
\begin{equation} \label{eq4}
    {{v}_{\text{remote}}}(t)=\cos ({{\omega }_{0}}t+{{\varphi }_{0}}+\Delta \varphi +{{\varphi }_{\text{DDS}}}),
\end{equation}
where $\Delta \varphi $ is the phase drift introduced by the signal passing through the optical fiber link. ${{v}_{\text{remote}}}(t)$ is connected to the remote FLOM-PD as the reference signal.

The output signal of the remote MLL passes through a $1\times 4$ fiber coupler, and the first optical signal is similar to the local site, attenuated to 10 dBm and then connected to the remote FLOM-PD to lock ${f_{\rm r}}$ of the remote MLL using the collaborative feedback control system. The second and third optical signals are respectively amplified after entering the channels C36 and C37 of the remote DWDM and then transmitted backward through the optical fiber link. Similarly, the \textit{n}-th harmonic frequency components of the two can be detected and expressed as
\begin{equation} \label{eq5}
\begin{split}
{{v}_{\text{C36}}}(t)=\cos (n{{\omega }_{0}}t+n{{\varphi }_{0}}+n\Delta \varphi +n{{\varphi }_{\text{DDS}}}), \\ 
{{v}_{\text{C37}}}(t)=\cos (n{{\omega }_{0}}t+n{{\varphi }_{0}}+n\Delta \varphi +n{{\varphi }_{\text{DDS}}}),
\end{split}
\end{equation}
where $n=1,2,3...$

The signal transmitted backward by the remote DWDM is amplified by the remote Bi-EDFA to about 3 dBm and then connected to the optical fiber link for return. After reaching the local site, the local Bi-EDFA amplifies the optical signal leaving the optical fiber link to about 3 dBm. Then, the amplified optical signal is connected to the local DWDM. Similar to the remote site, PD and BPF are used to filter out the first harmonic frequency component ${{v}_{\text{rt}}}(t)$ of the optical signal in channel C37. ${{v}_{\text{rt}}}(t)$ can be expressed as
\begin{equation} \label{eq6}
{{v}_{\text{rt}}}(t)=\cos ({{\omega }_{0}}t+{{\varphi }_{0}}+2\Delta \varphi +{{\varphi }_{\text{DDS}}}),
\end{equation}
where the asymmetry of the optical fiber link is ignored, which means the phase drift introduced by the optical fiber link is independent of the transmission direction.

The phase information 2$\Delta \varphi $ is obtained by using the local LIA to phase detect ${{v}_{\text{rt}}}(t)$ and ${{v}_{\text{local}}}(t)$. The phase information output by LIA is a voltage signal related to the amplitude, which a DMM collects and then reads and processes by a computer to obtain the phase information $\Delta \varphi $. The upper computer controls the DDS to shift the phase, that is
\begin{equation} \label{eq7}
{{\varphi }_{\text{DDS}}}=-\Delta \varphi ,
\end{equation}
then Eq. (\ref{eq4}) can be rewritten as
\begin{equation} \label{eq8}
{{v}_{\text{remote}}}(t)=\cos ({{\omega }_{0}}t+{{\varphi }_{0}}+\Delta \varphi -\Delta \varphi )=\cos ({{\omega }_{0}}t+{{\varphi }_{0}}),
\end{equation}
that means the frequency signal is restored at the remote site.

\subsection{DWDM-based node downloading.}\label{2.4}
Compared with the node downloading in \cite{grosche2008verfahren,gao2012fiber,krehlik2013multipoint}, the technical advantage of using MLL for node downloading is that the electrical signal obtained by the pulse output of MLL through PD contains rich frequency component information, so common electrical devices can be used to realize multiple harmonic frequency download output. Theoretically, frequency signals with frequency components of 200 MHz, 400 MHz, 600 MHz, and other 2\textit{N} times ${f_{\rm r}}$ can be recovered with the same frequency instability. In principle, the node frequency downloading can be achieved at any position of the link.

In this scheme, we exploit the wide spectrum characteristic of the pulse output of the MLL and further improve the node downloading structure. We do not use a $2\times 2$ fiber coupler at the node as in the previous scheme, but use a DWDM structure, which is a prevalent experimental structure in the optical fiber network node. By using the optical signals transmitted in the additional two channels C31 and C36, which are different from the active pre-compensation system for node downloading, we not only significantly reduce the interference to the active pre-compensation system but also can add or delete the node downloading structure at any time when the active pre-compensation system is working.

The output of the \textit{n}-th harmonic frequency component of the optical signal in channel C31 at the node after being detected can be expressed as
\begin{equation} \label{ep9}
{v}_{1}(t)=\cos (n{{\omega }_{0}}t+n{{\varphi }_{0}}+n{{\varphi }_{\text{DDS}}}+n{{\varphi }_{1}}),
\end{equation}
where $n=1,2,3...$ ${{\varphi }_{1}}$ is the phase drift introduced by the 50 km SMF from the local site to the node of the first harmonic frequency component. The \textit{n}-th harmonic frequency component of the optical signal in channel C36 after being detected can be expressed as
\begin{equation} \label{ep10}
{{v}_{2}}(t)=\cos (n{{\omega }_{0}}t+n{{\varphi }_{0}}+n{{\varphi }_{2}}+n{{\varphi }_{\text{DDS}}}+n\Delta \varphi ),
\end{equation}
where $n=1,2,3...$ ${{\varphi }_{2}}$ is the phase drift introduced by the 50 km SMF and 10 km DCF from the node to the remote site of the first harmonic frequency component, it is easy to know
\begin{equation} \label{ep11}
{{\varphi }_{1}}+{{\varphi }_{2}}=\Delta \varphi .
\end{equation}

The optical signal in channel C31 is detected by a PD, and the \textit{N}-th harmonic frequency component ${{v}_{3}}(t)$ is filtered out by a BPF with a center frequency of \textit{N} times ${f_{\rm r}}$. Through Eq. (\ref{eq7}), Eq. (\ref{ep9}) and Eq. (\ref{ep11}), ${{v}_{3}}(t)$ can be expressed as:
\begin{equation} \label{ep12}
{{v}_{3}}(t)=\cos (N{{\omega }_{0}}t+N{{\varphi }_{0}}-N{{\varphi }_{2}}).
\end{equation}

Similarly, the \textit{N}-th harmonic frequency component ${{v}_{4}}(t)$ in channel C36 can be expressed as:
\begin{equation} \label{ep13}
{{v}_{4}}(t)=\cos (N{{\omega }_{0}}t+N{{\varphi }_{0}}+N{{\varphi }_{2}}),
\end{equation}
where ${{v}_{3}}(t)$ and ${{v}_{4}}(t)$ are connected to a mixer after passing through amplifiers separately, and the high-frequency part ${{v}_{\text{note}}}(t)$ of the mixer output is obtained by using a BPF with a center frequency of 2\textit{N} times ${f_{\rm r}}$, that is:
\begin{equation} \label{ep14}
{{v}_{\text{note}}}(t)=\cos (2N{{\omega }_{0}}t+2N{{\varphi }_{0}}),
\end{equation}
where ${{v}_{\text{note}}}(t)$ is the frequency signal recovered at the node.

According to the analysis, the DWDM-based node downloading structure based on MLLs requires only a mixer, three BPFs, and two amplifiers. Moreover, the entire structure does not interfere with active pre-compensation, thus providing advantages such as simplicity, low cost, high practicality, and easy installation, maintenance, and update.

\section{Experimental results and analysis}\label{3}
\subsection{Performance and analysis of collaborative feedback control system.}\label{3.1} 
\begin{figure}[ht!]
\centering
\subfigure[]{\includegraphics[width=6.3cm]{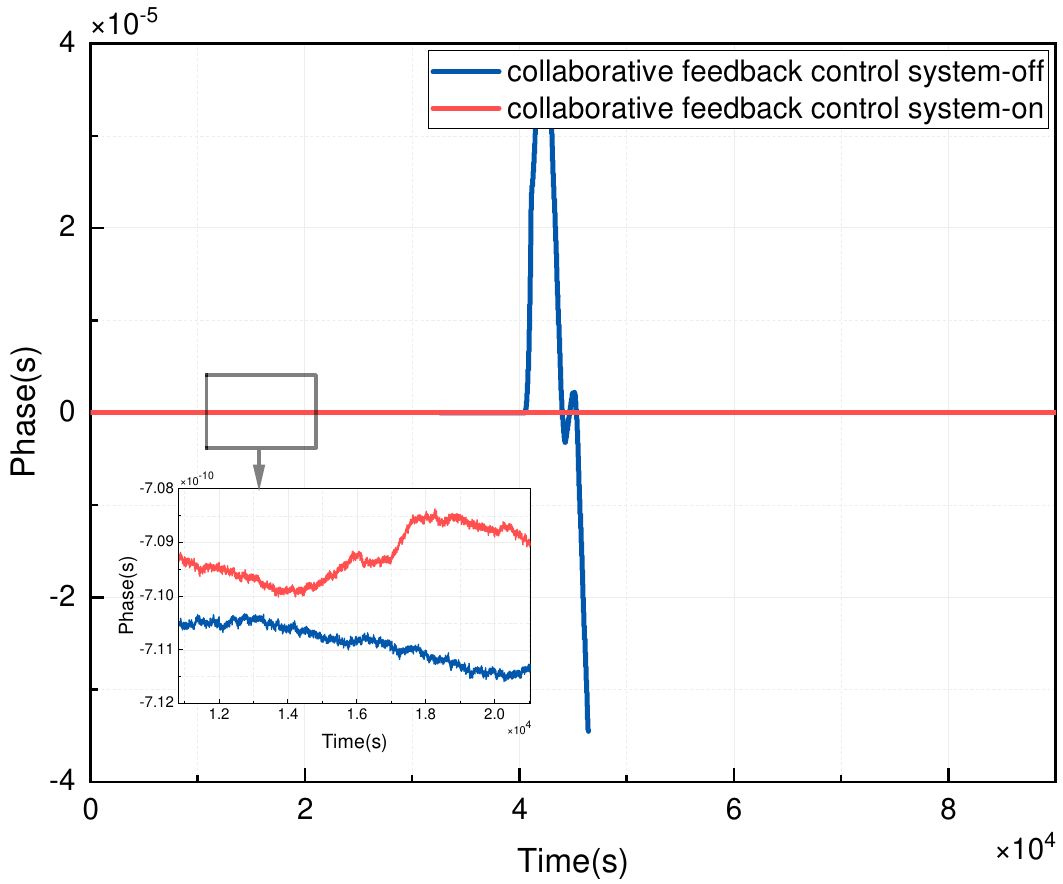}}
\quad
\subfigure[]{\includegraphics[width=6.5cm]{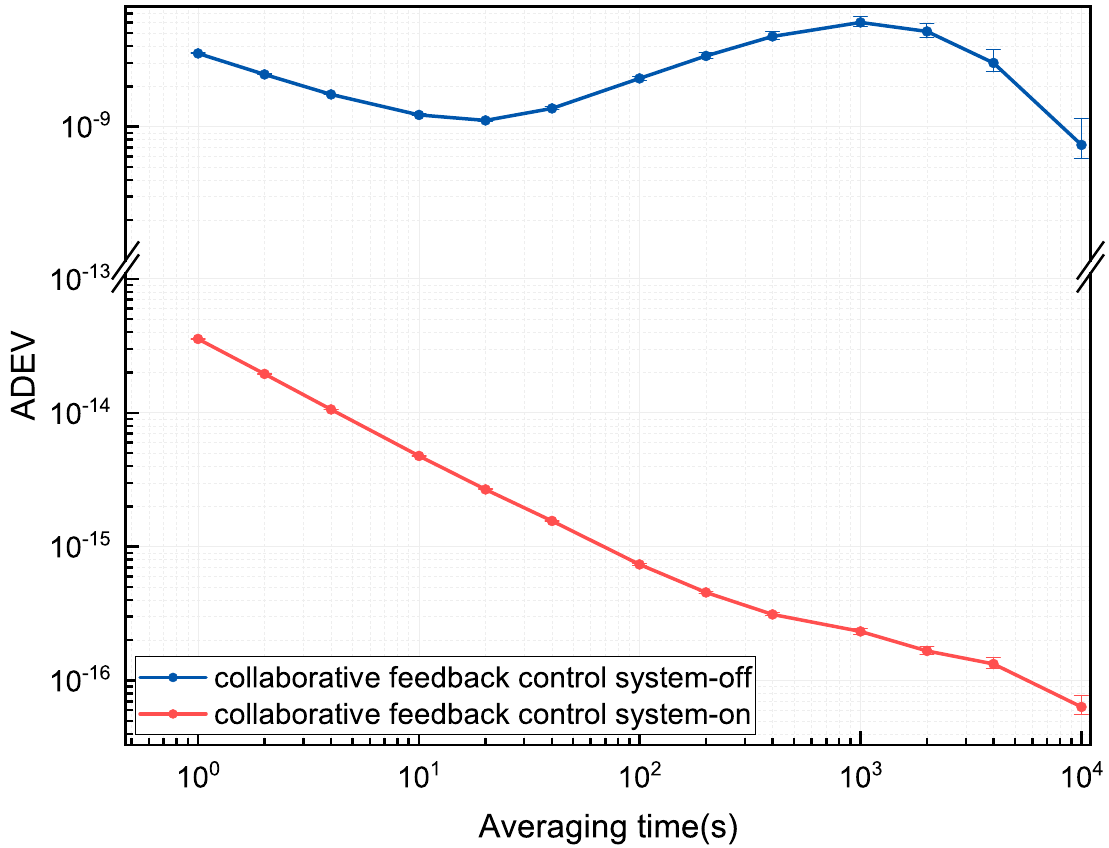}}
\caption{Result of the collaborative feedback control system. (a) The phase drift of the first harmonic of the mode-locked laser; (b) Allan deviations of the first harmonic of the mode-locked laser. The red line: collaborative feedback control system activated; the blue line: collaborative feedback control system deactivated.}
\label{result1}
\end{figure}

First, in order to verify the feasibility of the collaborative feedback control system, we turned off one of the air conditioners in the laboratory. We performed a long-term frequency locking of the MLL in an environment where the temperature difference between day and night was over 5 °C, and the ambient temperature fluctuated sharply. We measured the phase difference between the output of the Rb clock and the first harmonic of the mode-locked laser using a phase noise analyzer (\textit{Microchip, 53100A}), and then we utilized Stable32 to calculate the Allan deviation. The evaluation method in the following text is similar to this. The results can be seen in FIG. \ref{result1}. At approximately 40,000 s, the MLL without a collaborative feedback control system quickly lost the ${f_{\rm r}}$ locking, and then the phase jittered violently. The frequency instability is extremely worsened by nearly five orders of magnitude. The unlocking of the ${f_{\rm r}}$ occurred because the system entered the night at 40,000 s, and the temperature changed drastically when the air conditioner was not fully turned on. The temperature control adjustment speed was too slow to maintain the temperature stable, which caused ${f_{\rm r}}$ to drift beyond the PZT controllable range, resulting in ${f_{\rm r}}$ locking loss.

This experiment shows that by using the collaborative feedback control system, it is possible to achieve highly long-term ${f_{\rm r}}$ locking of the MLL under the condition that the temperature control and PZT tunable range of the MLL are not extensive, which reduces the cost and improves the versatility and practicality of the MLL. 

The experiment also demonstrates the importance of ${f_{\rm r}}$ locking of the MLL. It can be seen that after losing ${f_{\rm r}}$ locking, the phase drift increased to the order of $1\times {{10}^{-5}}$ s, at which level phase drift compensation is practically infeasible. Therefore, the long-term locking of the MLL plays an important role in the general stability of the system.

\subsection{Performance and analysis of active pre-compensation system.}\label{3.2}
Before evaluating the actual performance of the node downloading, it is important to assess the performance of the active pre-compensation. The frequency instability of the recovered frequency signal is evaluated at the remote site.

\begin{figure}[ht!]
\centering
\subfigure[]{\includegraphics[width=6.3cm]{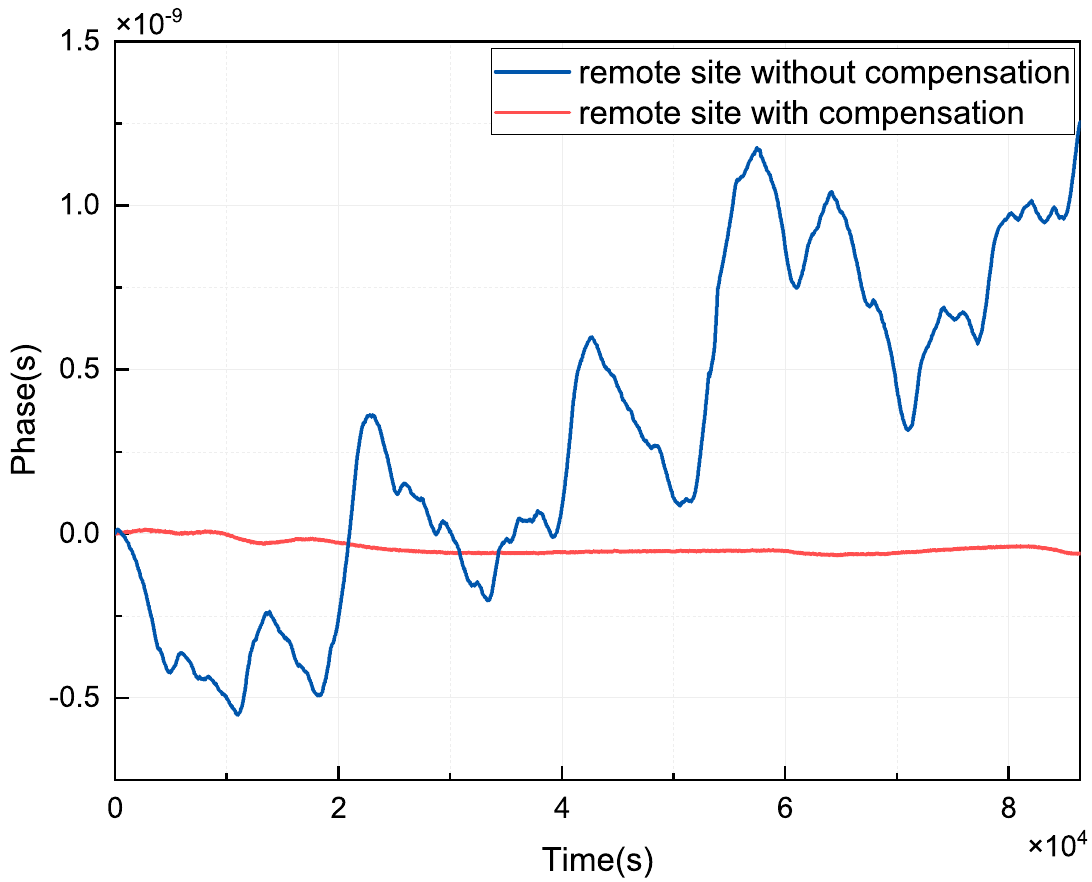}}
\quad
\subfigure[]{\includegraphics[width=6.3cm]{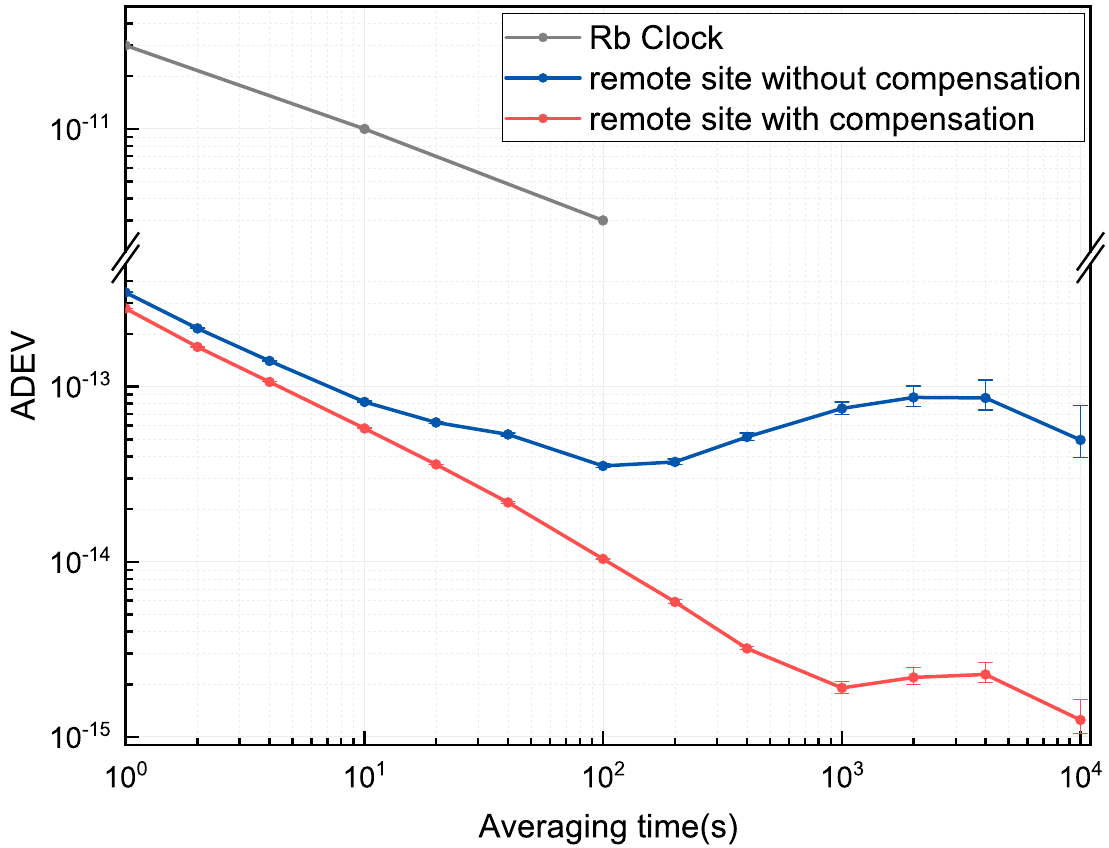}}
\caption{Result of the frequency transfer. (a) Phase drift of the frequency signal at the remote site; (b) Allan deviations of the Rb clock (gray line), the compensated (red line) and uncompensated (blue line) frequency transfers over 100-km fiber link.}
\label{result2}
\end{figure}

The frequency signal evaluated at the remote site is obtained after the fourth optical signal output by the $1\times 4$ coupler at the remote site passes through the PD and the BPF with a center frequency of 100 MHz. The evaluation results with and without the active pre-compensation are shown in FIG. \ref{result2}.

It can be seen that the active pre-compensation achieves frequency instabilities of $2.79\times {{10}^{-13}}$@1 s and $1.25\times {{10}^{-15}}$@10,000 s, which are slightly optimized for short-term instability compared to the instability when the phase drift was uncompensated, and optimized by more than one order of magnitude for long-term instability. The phase drift diagram also indicates that most of the significant phase changes have been eliminated after activating the active pre-compensation program. We note that the current results are sufficient to support the existing Rb clock transmission and better transmission results are expected to be achieved based on a clock with better performance as a reference, such as a state-of-the-art hydrogen clock \cite{marra2013transfer}.

\subsection{Performance and analysis of DWDM-based node downloading.}\label{3.3}
\subsubsection{Frequency instability of DWDM-based node downloading}
\begin{figure}[ht!]
\centering
\subfigure[]{\includegraphics[width=6.3cm]{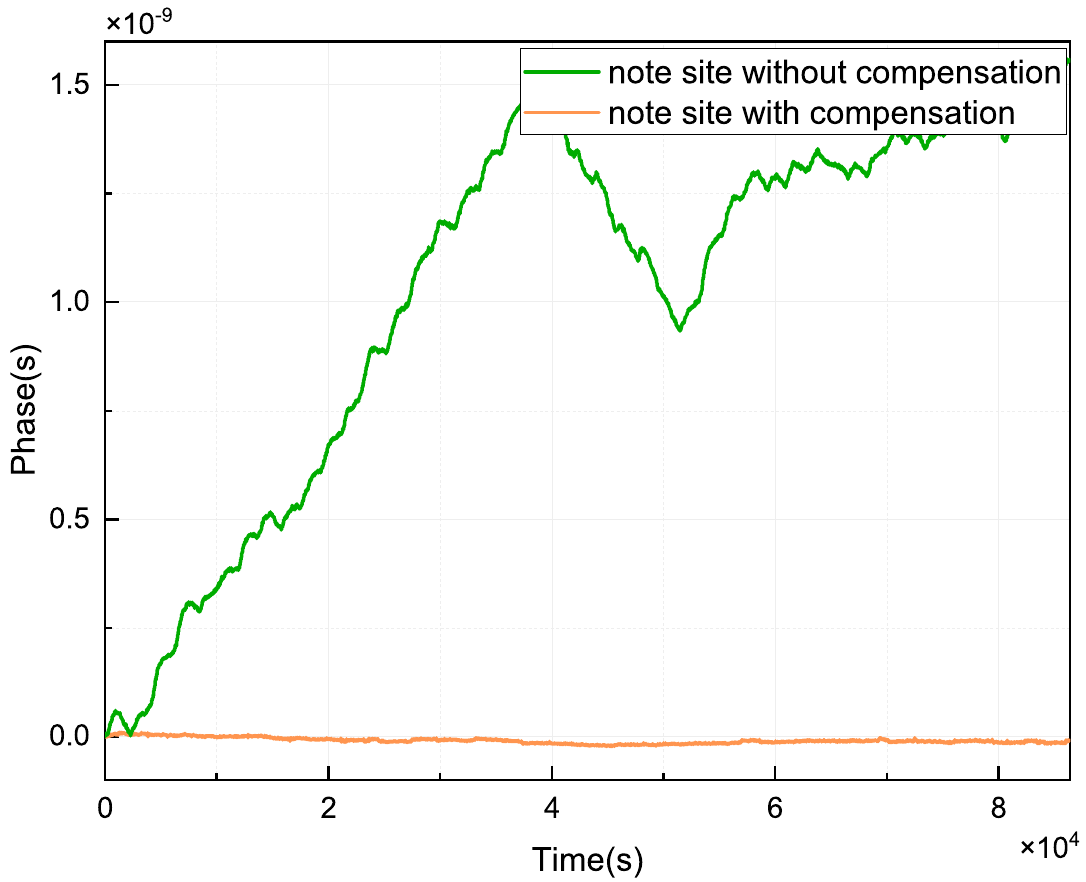}}
\quad
\subfigure[]{\includegraphics[width=6.3cm]{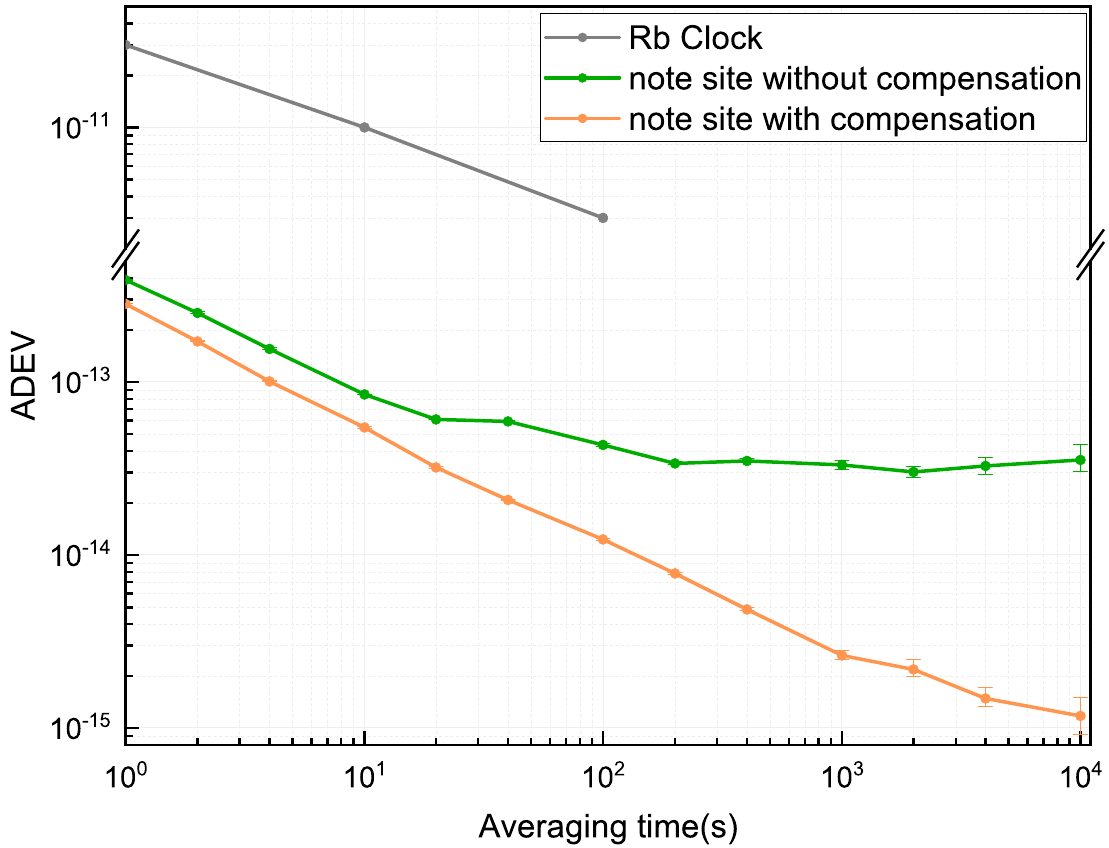}}
\caption{Result of the node downloading. (a) Phase drift of the frequency signal at the node; (b) Allan deviations of the Rb clock (gray line), the compensated (orange line) and uncompensated (green line) frequency downloading at the node.}
\label{result3}
\end{figure}

Under the optimized experimental parameters, we tested \textit{N}=1, which means using the first harmonic frequency component in the output of the MLL to obtain the reference signal ${{f}_{\text{node}}}$(t) at the node. We obtained the results of the short-term and long-term instability of the DWDM-based node downloading, $2.83\times {{10}^{-13}}$@1 s and $1.18\times {{10}^{-15}}$@10,000 s, as shown in FIG. \ref{result3}. It can be seen that compared with the evaluation results when the phase drift was uncompensated, DWDM-based node downloading and active pre-compensation are similar, with a slight optimization for short-term instability and a good optimization for long-term instability. Furthermore, the frequency instabilities of DWDM-based node downloading and active pre-compensation are not much different, especially for short-term instability. They achieve almost consistent performance. Similar performance implies that conducting node downloading at the node within the link of the frequency transfer system will not significantly degrade the frequency stability of the reference signal, which sufficiently satisfies the requirements of microwave frequency dissemination. As for the long-term stabilities, they show different behavior between node and remote sites, which was mainly due to temperature differences in the laboratory during testing. When testing at different times, it is difficult to ensure that the temperature conditions of the tested clock are precisely the same, which also further guides that in order to ensure consistent frequency stability at nodes and remote sites, we need to control laboratory environmental temperatures at different sites strictly.

In addition, we also analyzed some factors that influence the system's short-term instability: 1) The mixer has a frequency leakage phenomenon\cite{manstretta2003second}. As a result, the 100 MHz frequency signal at the RF port will leak to the LO port and then interfere with the output signal. The phase of that signal is inconsistent with the phase we aim to obtain, which will affect the frequency signal; 2) The electrical amplifier employed will cause additional electrical noise\cite{nguyen2004cmos}. We will complete further optimization of this part in our future research.

\subsubsection{Influence of different frequency harmonic components on the recovery of frequency signals in DWDM-based node downloading.} 
\begin{figure}[ht!]
\centering
\subfigure[]{\includegraphics[width=6.3cm]{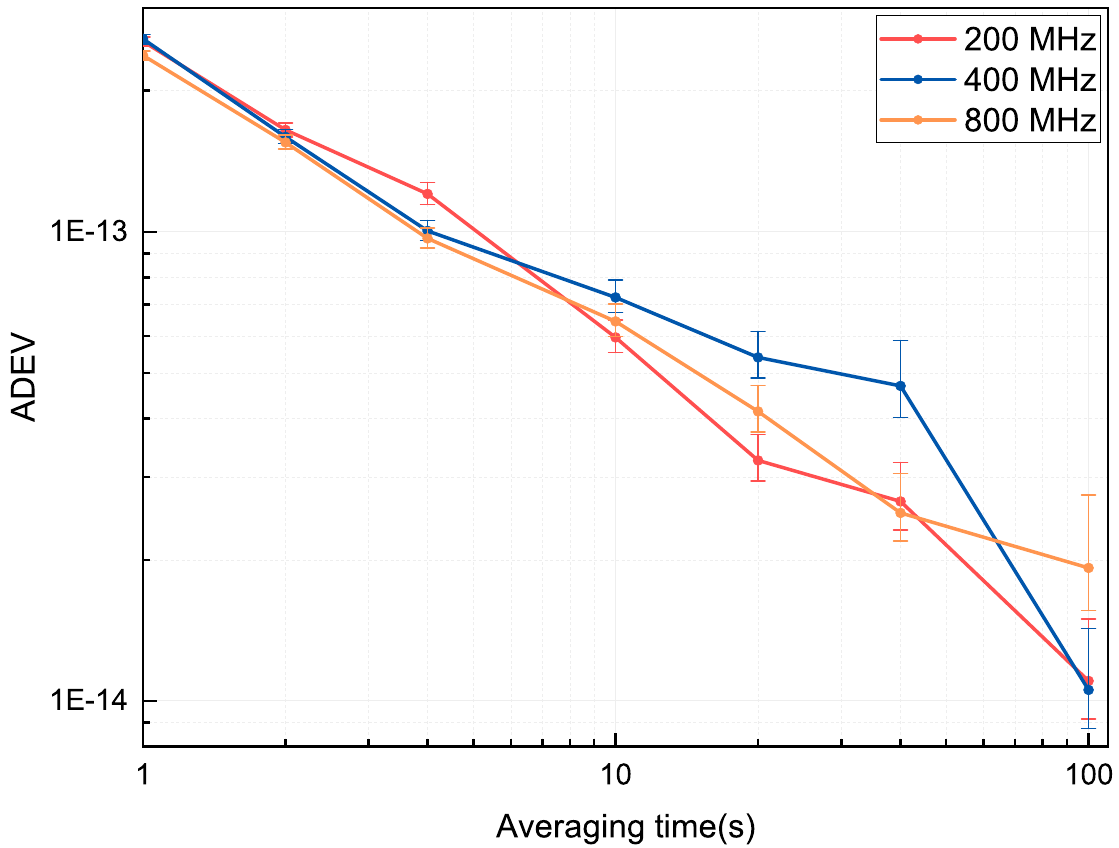}}
\quad
\subfigure[]{\includegraphics[width=6.3cm]{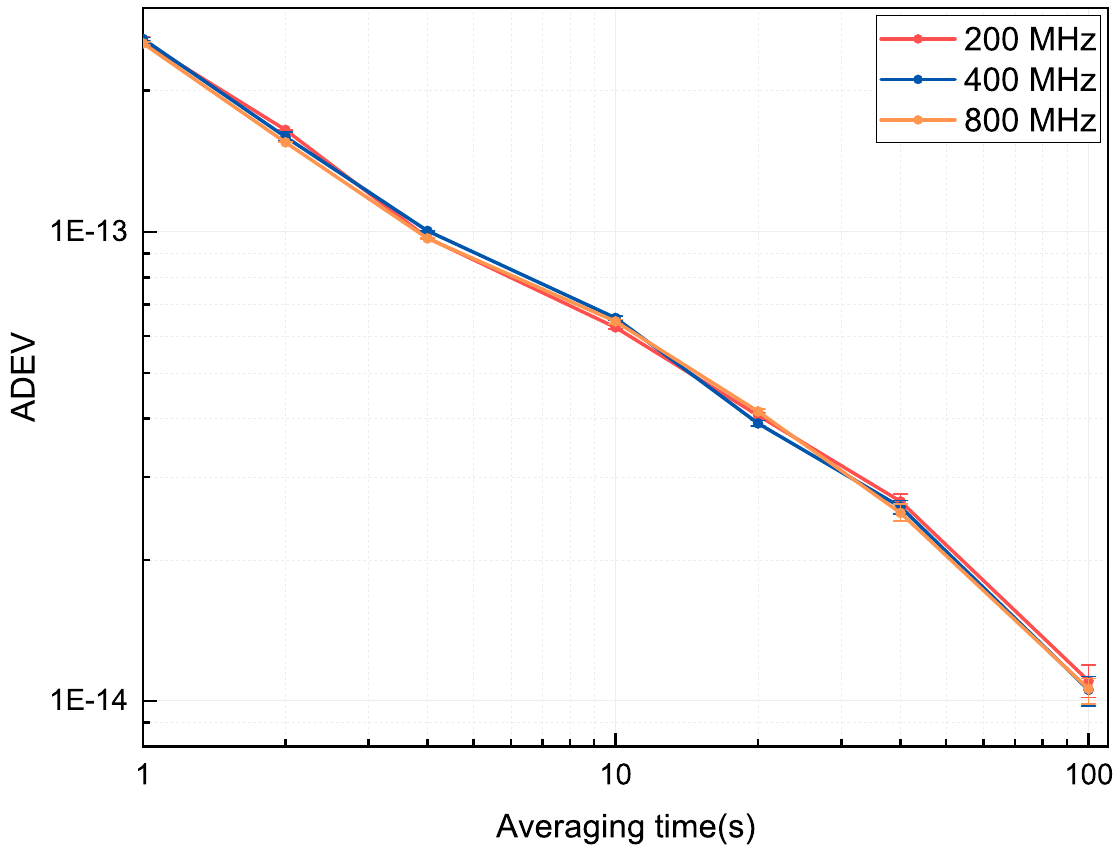}}
\caption{Allan deviations of the node downloading with different frequency. (a) 1200 data points without strict temperature control; (b) 5000 data points with strict temperature control. Red line: the frequency of the recovered signal at the node is 200 MHz; blue line: 400 MHz; orange line: 800 MHz. }
\label{result4}
\end{figure}

As can be seen from the principle part, theoretically, the reference signal with a frequency of 2\textit{N} times ${f_{\rm r}}$ can be recovered at the node by using the \textit{N}-th harmonic frequency component. We evaluated the frequency signals recovered by using 100 MHz, 200 MHz, and 400 MHz frequency components. Consequently, the frequencies of the recovered frequency signals are 200 MHz, 400 MHz, and 800 MHz, respectively, and the specific experimental results are shown in FIG. \ref{result4}. Note that the long-term stability of a system is primarily influenced by changes in environmental temperatures, and it has been deemed feasible to present only the performance characteristics of the system itself. Therefore, only the frequency instabilities for a limited duration of up to 100 seconds are given.

In this case, due to the insufficient data and the lack of strict temperature control, the consistency of the frequency components in FIG. \ref{result4} (a) is poor, and some frequency instability results exceed the error range. Therefore, we increased the data points from 1200 to 5000 and performed more rigorous temperature control. The results are shown in FIG. \ref{result4} (b). It is obvious that FIG. \ref{result4} (b) reflects the characteristics of the system more realistically. We encountered similar problems in the subsequent experiments. In order to accurately describe the performance of the system, we also performed data expansion and temperature control operations in the following manuscript.

\begin{figure}[ht!]
\centering\includegraphics[width=7cm]{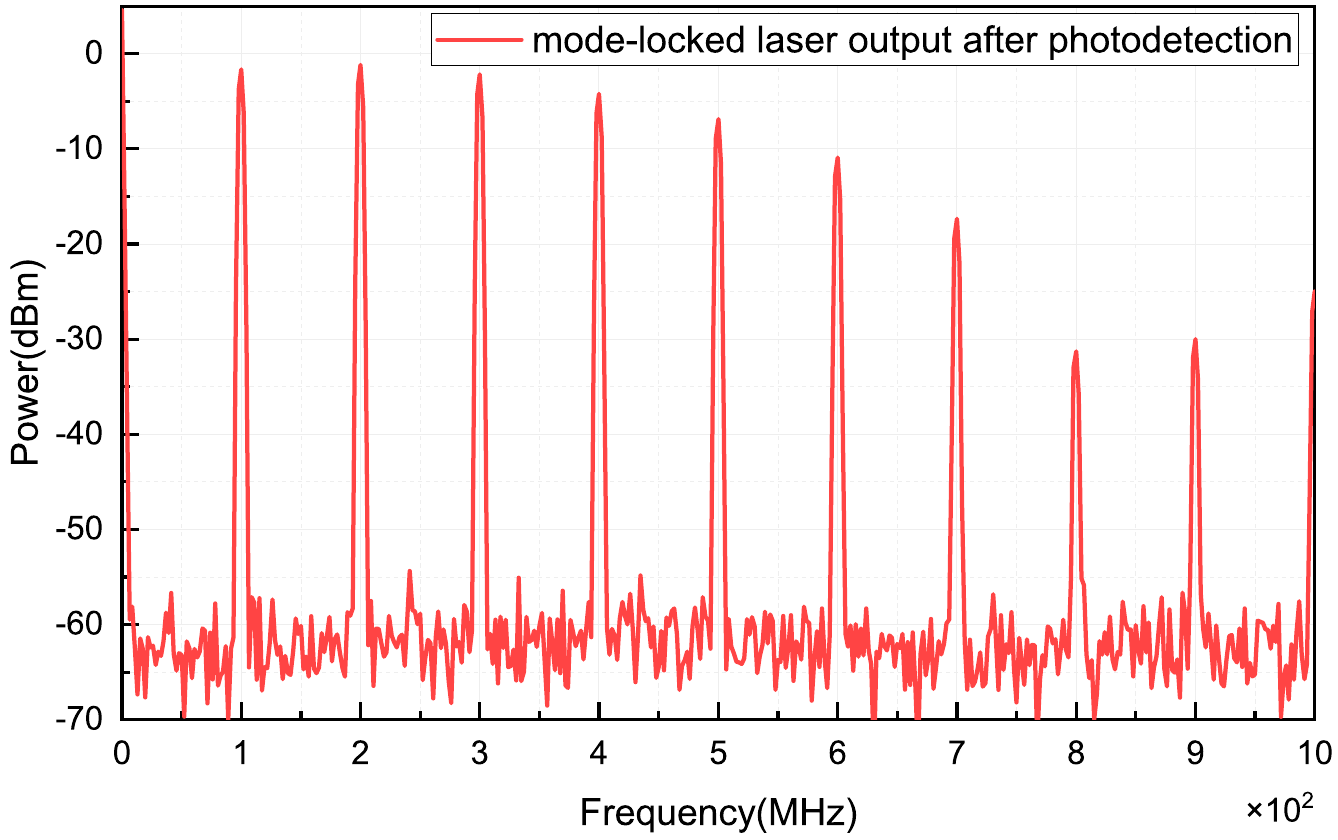}
\caption{Spectrum of the electrical signal after photoelectric conversion of the mode-locked laser signal.}
\label{result5}
\end{figure}

It can be seen that as \textit{N} increases, the short-term instability of the obtained reference signal almost remains constant, which fully meets the microwave frequency transfer requirements. This implies that if different frequency signals are needed at the node, it can be achieved by changing the BPF with a different center frequency and the mixer with a different working frequency without using complex electrical devices such as frequency doublers or frequency dividers, which greatly reduces the cost and enhances the practicality. However, we also noticed that as \textit{N} increases, the power of the obtained reference signal decreases. This is because the MLL pulse has a limited number of "teeth," so the higher harmonics have less power than the lower harmonics. We used a spectrum analyzer to analyze the output of direct photoelectric detection of the MLL pulse, as shown in FIG. \ref{result5}. It can be seen that after photoelectric conversion, the energy of the signal is mainly concentrated in the low-frequency part. To guarantee the usability of the reference signal at the node, the signal power should be maintained above a certain threshold. Moreover, if the reference signal power is too low, the phase detection accuracy of the mixer will also decrease gradually until it fails to evaluate normally. This is also why we did not use higher harmonics above 400 MHz in the MLL pulse to recover the reference signal at the node. For analyzing frequency signals beyond 800 MHz, it is possible to use amplifiers with increased gain, band-pass filters with reduced loss, and photodetectors with enhanced responsivity to amplify the signal power.

\subsubsection{Influence of channel selection on DWDM-based node downloading.}
\begin{figure}[ht!]
\centering
\subfigure[]{\includegraphics[width=6.3cm]{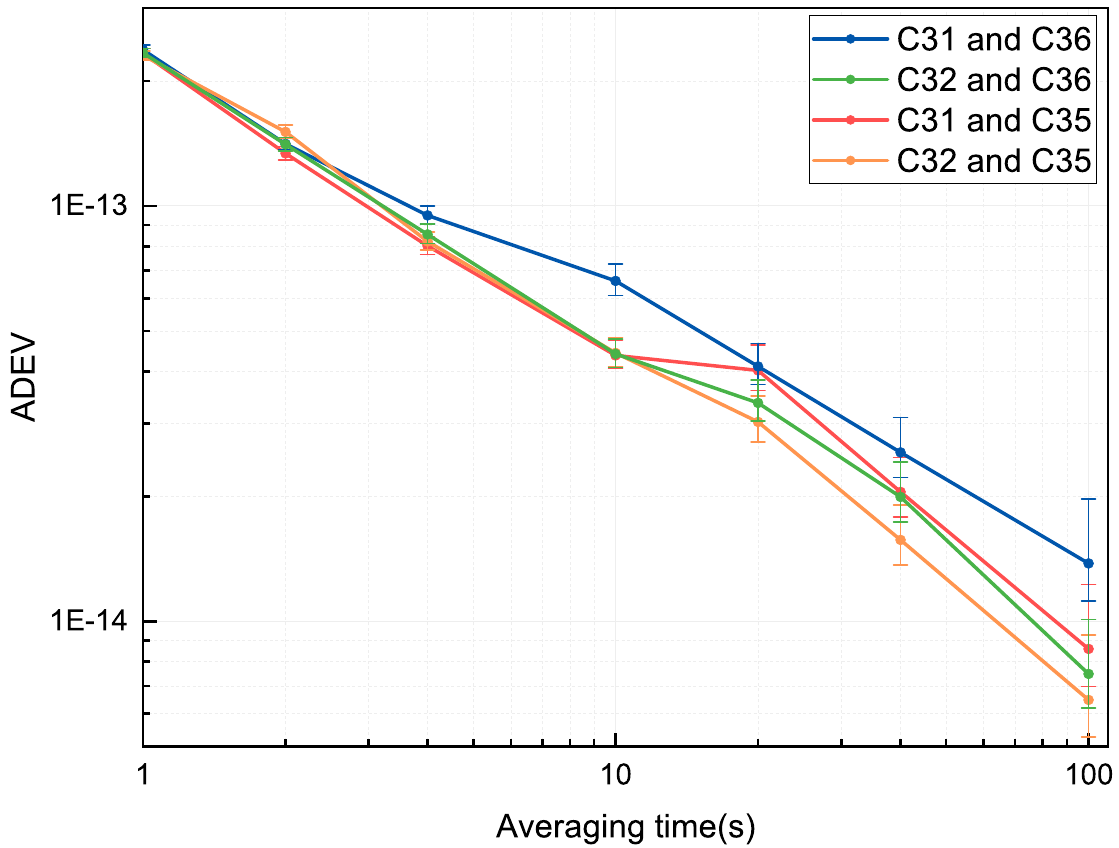}}
\quad
\subfigure[]{\includegraphics[width=6.3cm]{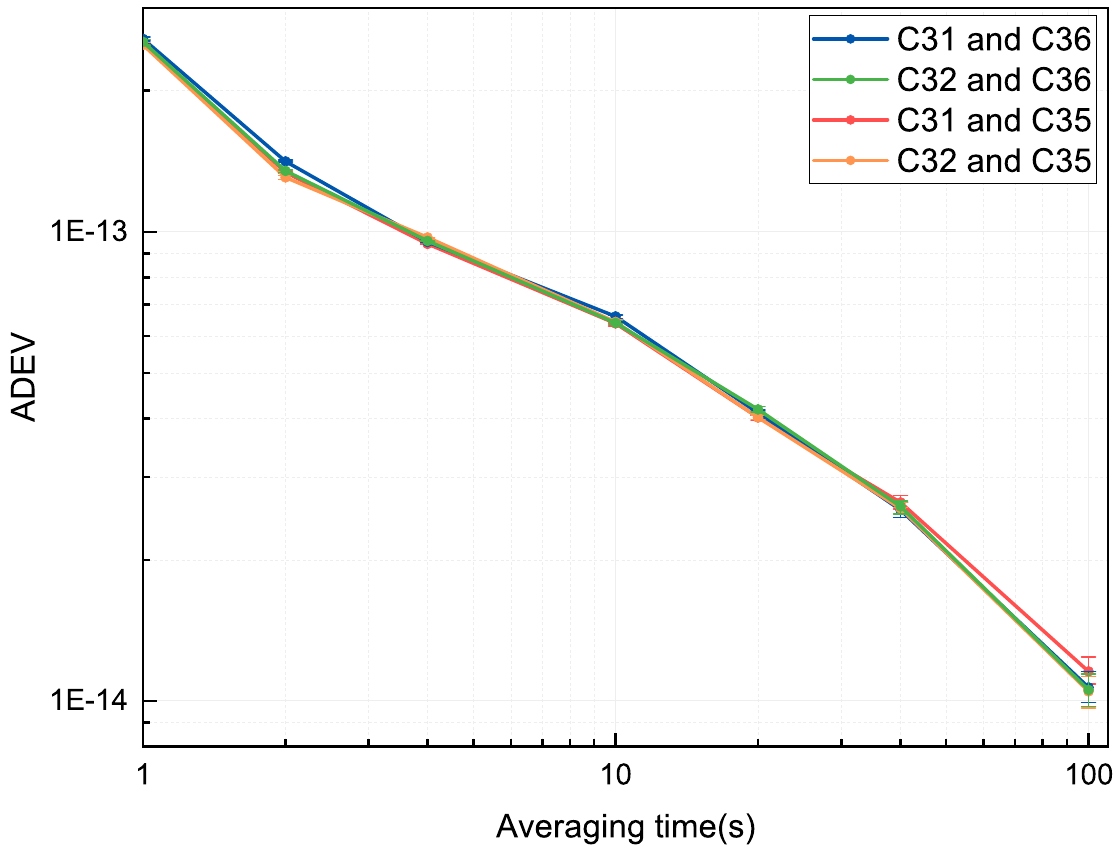}}
\caption{Allan deviations of the node downloading using different channel optical signals. (a) 1200 data points without strict temperature control; (b) 5000 data points with strict temperature control. Blue line: The channel used in the node downloading is C31 and C36; green line: C32 and C36; red line: C31 and C35; orange line: C32 and C35.}
\label{result6}
\end{figure}

The channels used for node downloading in FIG. \ref{Experimental} are C31 and C36. As previously mentioned, the spectrum of the MLL pulse is wide. It covers each channel of the DWDM we use with high energy, enabling other channels to be used for node downloading as well. This facilitates the addition of multiple node downloading structures in a frequency transfer system. We choose different channels except C30 and C37 for node downloading, and the results are shown in FIG. \ref{result6}. It can be seen that there is almost no difference in the results obtained by using any pair of channels among C31 and C36, C32 and C36, C31 and C35, C32 and C35 for node downloading. The channel selection has a negligible influence on node downloading. This means that any two channels other than the ones involved in the active pre-compensation can be used for node downloading at the node.

At the same time, We observe that the frequency instability is not affected by the forward and backward transmission signals being in the same channel in the DWDM-based node downloading structure because they are cut off after reaching the node by the addition of this structure. Thus, there is no cross-transmission phenomenon. This means that any channel except for the ones involved in active pre-compensation can be used for node downloading at any node. However, in our experiment, the Bi-EDFA at the local and remote sites has a built-in DWDM with channels C26-C33 for forward working wavelength and channels C34-C41 for backward working wavelength. The forward and backward transmission signals cannot share the same channel. We plan to improve the Bi-EDFA structure to amplify both signals in a single channel. This will increase the number of DWDM-based node downloading structures that can be added to the frequency transfer system by twofold.

\begin{figure}[ht!]
\centering
\subfigure[]{\includegraphics[width=6.3cm]{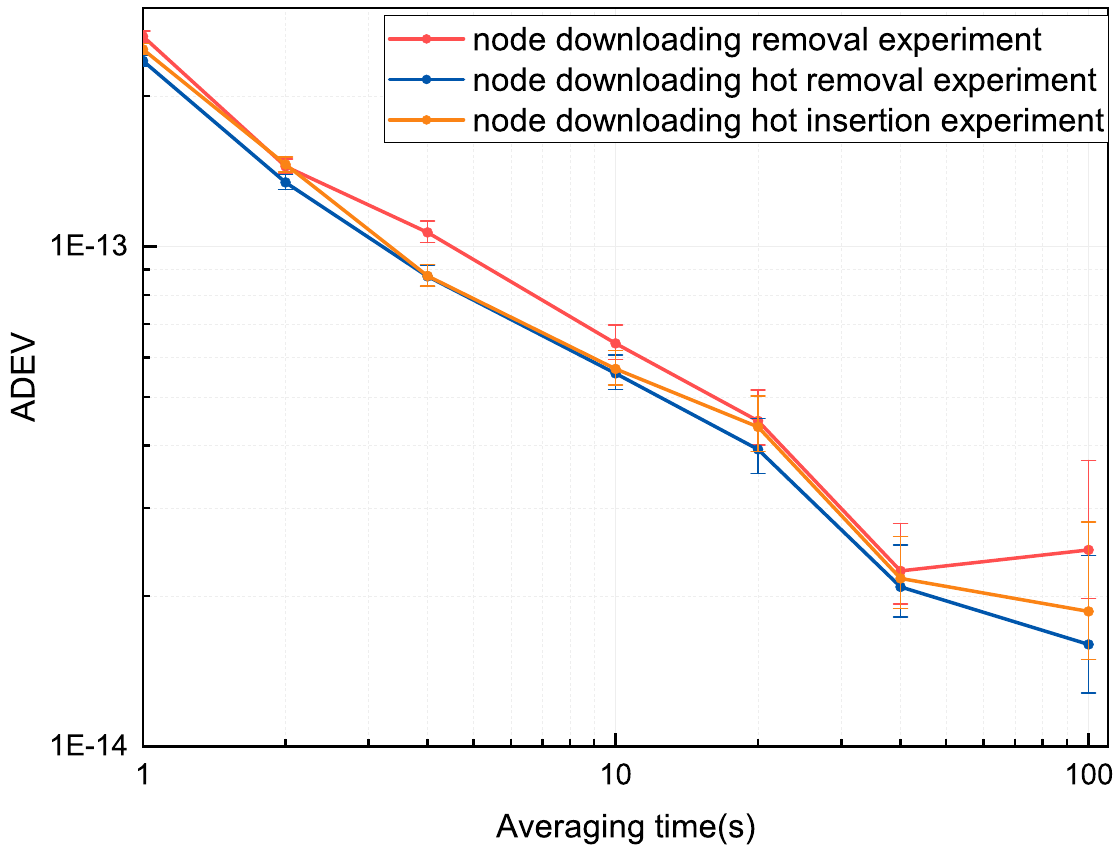}}
\quad
\subfigure[]{\includegraphics[width=6.3cm]{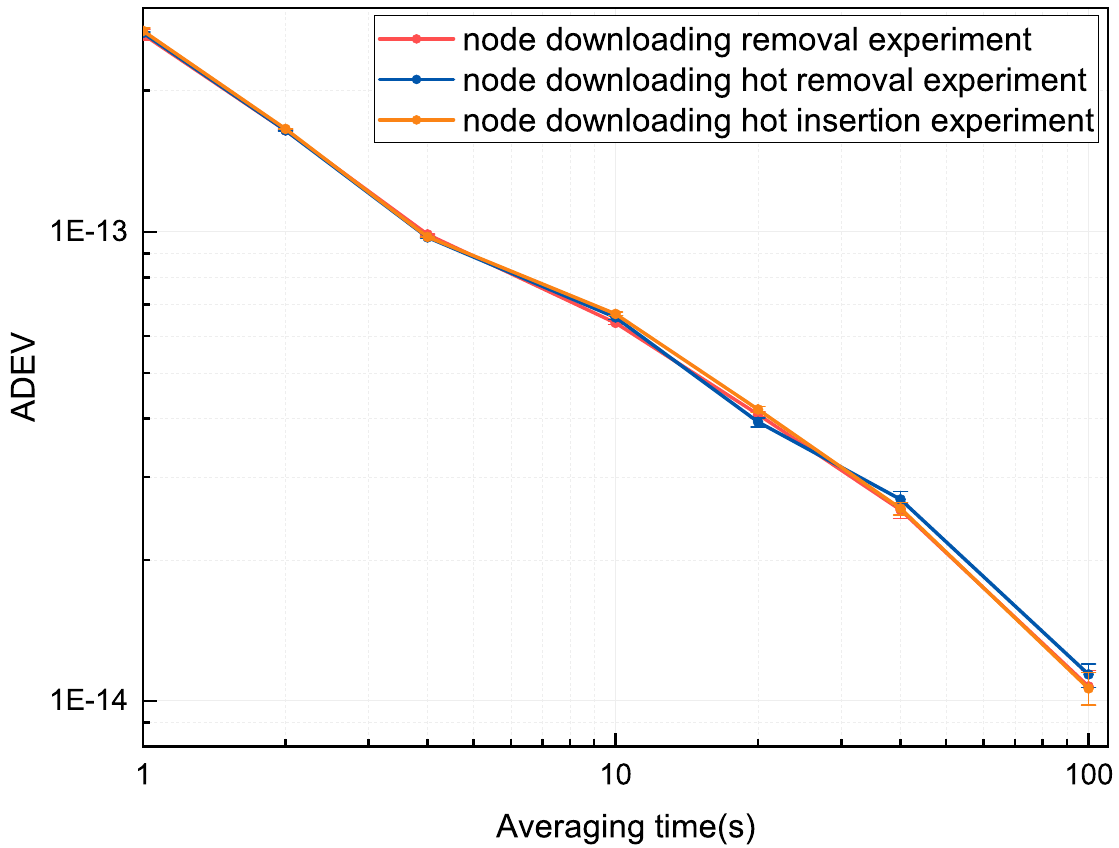}}
\caption{Allan deviations of the node downloading in the plug-and-play tests. (a) 1200 data points without strict temperature control; (b) 5000 data points with strict temperature control. Red line: Node downloading removal experiment; blue line: Node downloading hot removal experiment; orange line: Node downloading hot insertion experiment.}
\label{result7}
\end{figure}

\subsubsection{Performance of plug-and-play node feature.}
To evaluate whether the DWDM-based node downloading has any impact on the frequency transfer system, we performed the following three experiments: 1) Node downloading removal experiment, in which we directly connected the channel C31 and C36 of the two-stage DWDM at the node with optical fibers, turned on the active pre-compensation system, and evaluated the reference signal at the remote site; 2) Node downloading hot removal experiment, in which we turned on the active pre-compensation system, evaluated the reference signal at the remote site, and removed the node downloading structure at the 60th second of the evaluation process; 3) Node downloading hot insertion experiment, in which we removed the node downloading structure, turned on the active pre-compensation system, evaluated the reference signal after the phase-locked loop, and added the node downloading structure at the 60th second of the evaluation process. The time for adding and removing the node downloading structure does not exceed 5 s. The results are shown in FIG. \ref{result7}. It is evident that there is almost no difference in performance among the three experiments, which indicates that the DWDM-based node downloading structure does not affect the performance of the active pre-compensation system. It can also be arbitrarily added or removed during the system operation.

Currently, the short-term instability of the reference signal at the remote site and the node is not well optimized relative to the uncompensated state, and periodic variations can still be observed in the phase drift. This is most likely because of the use of DDS, which has a time difference of about 200 $\upmu$s between each compensation, and the compensation rate limits the improvement of short-term instability. At the same time, the LIA and the host computer program that we use will also cause accuracy loss, so the phase drift is not fully compensated. In the future, if we can reduce the delay between the phase shifts of DDS and improve the resolution of LIA, we expect that both the short-term and long-term instability of active pre-compensation and node downloading can be further optimized.

\section{Conclusion}\label{4}
We demonstrated a node-downloadable frequency transfer system using MLL technology via 100 km of spooled fiber. In the system, we used active pre-compensation to eliminate the phase drift introduced by the fiber link. We realized the frequency transfer with frequency instabilities of $2.79\times {{10}^{-13}}$@1 s and $1.25\times {{10}^{-15}}$@10,000 s at the remote site. At the node 50 km away from the local site, we recovered several frequencies with frequency instabilities of $2.83\times {{10}^{-13}}$@1 s and $1.18\times {{10}^{-15}}$@10,000 s, which reached the similar instabilities of the remote site.

Since we exploit the characteristic of MLL containing multiple frequency components, this not only avoids the increase of node structure complexity and performance degradation caused by the use of electrical devices such as frequency multipliers, dividers, etc., but also supports the simultaneous downloading of multiple frequency components at the node, which expands the frequency applications at the node. Based on these technologies, it is expected that stable multi-frequency components can be obtained from any location on the link in the time-frequency networks in the future, providing technical support for the frequency use of each node in long-distance frequency transfer.

\begin{acknowledgments}
	This work was supported by the National Natural Science Foundation of China (Grants Nos. 62201012, and 61531003) and National Hi-Tech Research and Development (863) Program.
\end{acknowledgments}

\end{document}